\documentclass[]{aa}

\usepackage{txfonts}
\usepackage{graphicx, natbib}
\bibpunct{(}{)}{;}{a}{}{,}

\begin{document}
\title{Optical spectropolarimetry with incomplete data sets}
\author{Justyn R. Maund} \offprints{Justyn Maund, \email{jrm@astro.as.utexas.edu}} 

\institute{Department of Astronomy and McDonald Observatory, The University
of Texas at Austin, 1 University Station, C1400, Austin, Texas
78712-0259, U.S.A.}
\date{Received /Accepted } 

\abstract{Linear spectropolarimetry is a ``photon-hungry'' observing
technique, requiring a specific sequence of observations to determine
the Stokes $Q$ and $U$ parameters.  For dual-beam spectropolarimeters,
the $Q$ and $U$ Stokes parameters can be ideally determined using
observations at N=2 retarder plate positions.  The additional
polarization signal introduced by instrumental effects requires the
redundancy of N=4 observations to correct for these effects and to
accurately measure the linear polarization of astronomical objects.}
{We wish to determine if the ``instrumental signature corrections''
for the Stokes $Q$ and $U$ parameters, $\epsilon_{Q}$ and
$\epsilon_{U}$, are identical for observations with dual-beam
spectropolarimeters.  For instances when observations were only
acquired at N=3 retarder plate angles, we wish to determine if the
complete measurement of one Stokes parameter and the associated
instrumental signature correction can be used to determine the other
Stokes parameter.}  {We constructed analytical and Monte Carlo models of
a general dual-beam spectropolarimeter to study the factors affecting
the assumption $\epsilon_{Q}=\epsilon_{U}$ and the uncertainty
thereon.  We compared these models with VLT FORS1 linear
spectropolarimetry observations.}  {We find that, in general,
$\epsilon_{Q}-\epsilon_{U}\approx0$, with the variance around zero
($\Delta(\epsilon_{Q}-\epsilon_{U})$) being directly related to the
signal-to-noise ratio of the observations.  Observations of a
polarized standard star, observed under identical instrumental
conditions over the period of 2002-2007, show that the assumption of
$\epsilon_{Q}-\epsilon_{U}=0$ is generally true over a long period,
although the absolute values of $\epsilon_{Q}$ and $\epsilon_{U}$ vary
between observational epochs.  While the variance of
$\epsilon_{Q}-\epsilon_{U}$ is not dependent on the polarization
angle, significant deviations from $\epsilon_{Q}-\epsilon_{U}=0$ arise
when $p \gtrsim 20\%$.}  {Incomplete VLT FORS1 spectropolarimetry
datasets, for which observations at only N=3 retarder plate position
angles have been acquired, can be analyzed under the assumption that
$\epsilon_{Q}\approx\epsilon_{U}$.  The uncertainty associated with
this assumption is directly related to the signal-to-noise ratio of
the observations.  This property of the analysis of
spectropolarimetry, with dual beam spectropolarimeters, can be used
to test for the presence of artifacts affecting individual
observations and to assess the quality of the data reduction, when
observations at all four retarder plate angles have been acquired.}

\keywords{instrumentation:polarimeters - techniques:polarimetric}
\maketitle

\section{Introduction}
\label{section:introduction}
Spectropolarimetry is a ``photon-hungry'' observational technique,
requiring a particular sequence of observations to completely and
independently determine the polarization properties of astronomical
objects.  In order to measure particularly low levels of polarization
($\lesssim 0.1\%$), high signal-to-noise ratio (S/N) observations are
required at each step in the observing sequence.\\ For dual-beam
spectropolarimeters, such as the European Southern Observatory (ESO)
Very Large Telescope (VLT) FOcal Reducer and low dispersion
Spectrograph (FORS1; \citealt{1998Msngr..94....1A}), the determination
of the linear polarization Stokes parameters, $Q$ and $U$, requires
N=4 separate observations.  This observing sequence introduces
redundancies, but permits the complete determination and correction
for instrumental effects that would, if uncorrected for N=2
observations, lead to spurious observed polarization.\\ Target of
Opportunity spectropolarimetry (where observations are conducted
at specific epochs to study transient events which occur without prior
warning) of time-variable phenomena is at a disadvantage, since at
each epoch the technique requires a factor $\sim 8$ more time on
target than for pure spectroscopy to achieve the same S/N in the
individual spectra.  For objects such as SNe, the position of the
target on the sky may not permit sufficient time on target to conduct
all of the necessary observations, leaving the dataset acquired
incomplete.  If one observation, for one retarder plate angle, is
absent, only one Stokes parameter can be independently measured.
There is, however, potential redundancy between the observations for
the completely determined Stokes parameter and the partially
determined parameter such that instrumental effects can be removed,
allowing for complete determination of both parameters, but at a
higher degree of uncertainty.\\ \citet{2001ApJ...553..861L} present a
technique where, by assuming the same ratio of gains for the ordinary
and extraordinary rays, instrumental effects can be removed to
calculate the second Stokes parameter for incomplete datasets.  When
``normalized flux differences''
\citep[e.g.][]{forsman,2006PASP..118..146P} are used to calculate the
Stokes parameter, the consideration is of a correction for the
instrumental signature, which quantifies the deviation of data from
the ideal case of N=2.\\ Here we present a technique to determine the
instrumental signature corrections for observed data, such that, in
the event of an incomplete observation, both Stokes parameters can be
determined.  The concept of spectropolarimetry with dual beam
spectropolarimeters is, briefly, outlined along with discussion of
correction for instrumental effects in \S\ref{section:basicconcept}.
In \S\ref{section:models}, analytical and Monte-Carlo models of a dual
beam spectropolarimeter are presented, and the effects of various
factors on the final measured polarization for complete and incomplete
datasets are presented.  Real observations of linearly polarized point
sources, acquired using VLT FORS1, and the instrumental signature
corrections are presented in \S\ref{section:observations}.\\

\section{Basic Concept and Theory}
\label{section:basicconcept}
Dual-beam spectropolarimeters, such as the FORS1 instrument
\citep{1998Msngr..94....1A}, use a sequence of a retarder plate and a
Wollaston prism to measure the polarization components.  
Rotation of the retarder plate varies the angle at which the
orthogonal polarization components are sampled.  The Wollaston prism
separates these two components spatially into the ordinary (${o}$) and
extraordinary rays (${e}$).  At a general retarder plate position $i$,
the {\it measured} normalized flux difference $F^{m}_{i}$ is defined \citep[e.g.][]{forsman} as
\begin{equation}
F^{m}_{i}=\frac
{f^{i}_{o}-f^{i}_{e}}
{f^{i}_{o}+f^{i}_{e}}
\label{forseqtn}
\end{equation}
such that the total intensity $I=f_{o}+f_{e}$.  $F^{m}_{i}$ is
normalized by the total flux intensity and is independent
of varying sky transparency and exposure times.\\ 
The optimum observing sequence is for
a half-wavelength retarder plate to be positioned at N=4 position angles
$\theta_{i}$: $\theta_{0}$=0\fdg0, $\theta_{1}$=22\fdg5,
$\theta_{2}$=45\fdg0 and $\theta_{3}$=67\fdg5.  

At $\theta_{0}$, the
orthogonal polarization components are in the horizontal and vertical
directions.  For $\theta_{2}$ the same components are observed, but
the beams are swapped such that $f^{i}_{o}=f^{i+2}_{e}$.  Similarly,
for $\theta_{1}$ and $\theta_{3}$ the diagonal polarization components
are observed, and the polarization components observed as the ordinary
and extraordinary rays are switched between the two observations.
The normalized Stokes parameters are given, in terms of normalized flux differences, by \citet{forsman} and
\citet{2006PASP..118..146P}, as
\begin{equation}
q= \frac{1}{2}\left( F^{m}_{0}-F^{m}_{2}\right) = \frac{1}{2}\sum_{i=0}^{3}
\left( F^{m}_{i}\right) \cos \left( \frac{\pi}{2} i \right)
\end{equation}
\label{qeqtn}
\begin{equation}
u=
\frac{1}{2}\left( F^{m}_{1}-F^{m}_{3}\right)
=
\frac{1}{2}\sum_{i=0}^{3} \left( F^{m}_{i}\right) \sin \left( \frac{\pi}{2} i \right)
\label{ueqtn}
\end{equation}
where the flux Stokes parameters are given by $Q=qI$ and
$U=uI$.  The redundancy in N=4 observations permits the removal of the
instrumental effects that differ between the {\it o} and {\it e} rays.  These differences
manifest themselves as spurious polarization or depolarization.  These
instrumental effects are discussed by \citet{2006PASP..118..146P}.
A gain difference between the {\it o} and
{\it e} rays for N=2 observations would manifest itself as
significant polarization.  For spectropolarimetry, flatfields are
acquired with the full polarization optics in place, such that the
observed flatfields, themselves produced by scattered light, are
polarized.  In addition, the optical components following the
analyzer, such as grisms, filters and lenses, can also act act as
linear polarizers, producing a constant additive polarization
component which is larger than effects due improper flatfielding using
unpolarized flats \citep{2006PASP..118..146P}.  Importantly,
\citeauthor{2006PASP..118..146P} also identify a non-additive
polarization term, which can arise from a non-ideal Wollaston prism;
for most modern dual-beam spectropolarimeters, such as FORS1, the
imperfections of the Wollaston prism and the associated effects are
negligible.  
In the case of N=3 observations, the instrumental polarization
component cannot be removed from the Stokes parameter for which there
was only a single observation.\\ At each retarder plate angle, the
measured value of the normalized flux difference can be considered as
the sum of the ideal normalized flux difference ($F_{i}$) and the instrumental
signature correction $\epsilon$: $F_{i}^{m}=F_{i}+\epsilon$, such that
under ideal conditions ($\epsilon=0$) a Stokes parameter can be
measured using only one value of $F$ at only one retarder plate
position (such that the $q$ and $u$ parameters are completely
determined with N=2 observations).\\ 
In the same
form as Eqns. \ref{qeqtn} and \ref{ueqtn}, the instrumental signature
corrections for the $q$ and $u$ Stokes parameters are,
therefore, given by
\begin{equation}
\epsilon_Q=
\frac{1}{2}\left( F^{m}_{0}+F^{m}_{2}\right)
=\frac{1}{2}\sum_{i=0}^{3}  F^{m}_{i} \cos^{2}\left( \frac{\pi}{2} i \right)
\label{eqeqtn}
\end{equation}
\begin{equation}
\epsilon_U=
\frac{1}{2}\left( F^{m}_{1}+F^{m}_{3}\right)
=\frac{1}{2}\sum_{i=0}^{3}  F^{m}_{i} \sin^{2}\left( \frac{\pi}{2} i \right).
\label{eueqtn}
\end{equation}
Another benefit of this formalism is that the instrumental signature
corrections are flux normalized (such that
$\epsilon_{Q}$ and $\epsilon_{U}$ are {\it percentages} of the total
flux) and are independent of the same factors as the
normalized flux differences.\\ In the observing sequence, the primary
change between each exposure is the rotation of the retarder plate.
For spectropolarimetry of a point source at the centre of the field, the only change
should be the orientation of the retarder plate, with the location of
the source on the retarder plate unchanged\footnote{If the
source is not at the centre of the field passing the beam through different points in the retarder plate, as
the retarder plate is rotated, may become important.}.  In this case, therefore, the values of the corrections for the $Q$ and $U$ Stokes parameters should be
approximately identical.  For an
observing sequence with N=3, the normalized Stokes parameter with incomplete observations  can be determined from the correction determined for the other {\it completely determined} parameter, assuming $\epsilon_{Q}=\epsilon_{U}$, as
\begin{equation}
q \approx F^{m}_{0}-\epsilon_{U} \approx -(F^{m}_{2}-\epsilon_{U})
\end{equation}
\begin{equation}
u \approx F^{m}_{1}-\epsilon_{Q} \approx -(F^{m}_{3}-\epsilon_{Q}).
\end{equation}
The change in sign of the instrumental signature correction and $F$ is due to switch of the polarization
components observed as the {\it o} and {\it e} rays.
\section{Models}
\label{section:models}
\subsection{Analytical Model}
\label{section:models:analytic}
\begin{figure}
\includegraphics[width=7cm, angle=270]{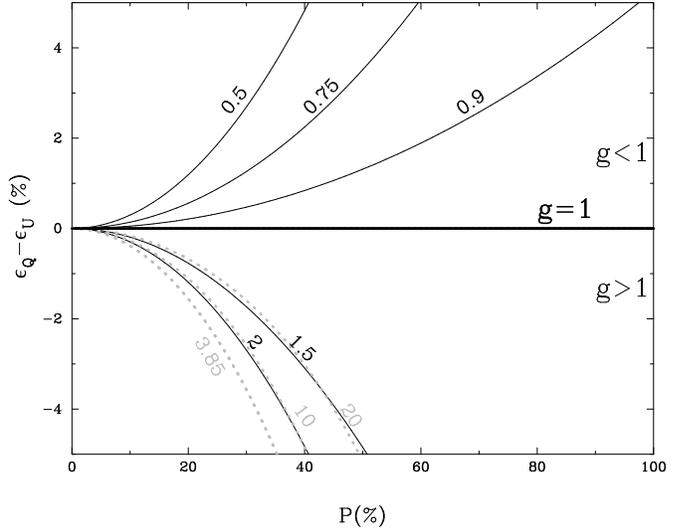}
\caption{The analytic form of the dependence of the instrumental
signature correction on the gain ratio between the $o$ and $e$ rays
and the total polarization, for the case of $p=q$ and $u=0$.  For the
opposite case, $p=u$ and $q=0$, the top and bottom regions of the
diagram are switched.  For $g>3.85$ $\epsilon_{Q}-\epsilon_{U}$ is
shown as grey dotted lines.}
\label{fig:model:anal} 
\end{figure}

\begin{figure}
\includegraphics[width=7cm, angle=270]{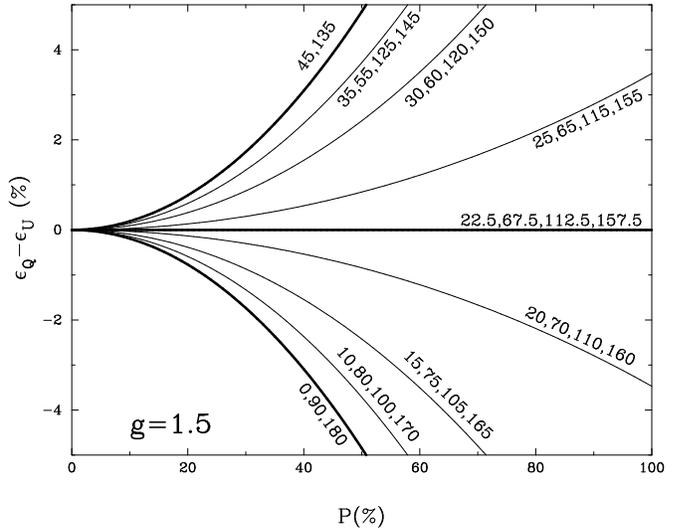}
\caption{The difference of the instrumental signature corrections for 
varying polarization angles, for a fixed gain ratio $g=1.5$.}
\label{fig:model:angle}
\end{figure}

If the principal difference between the measured $o$ and $e$ rays is purely due
to sensitivity, such that the difference can be expressed as a ratio of
the gains between the {\it o} and {\it e} rays $g_{o}/g_{e}=g$, then in terms of the {\it ideal} values of $f_{o}$ and $f_{e}$ (with no instrumental effects) Eqn. \ref{forseqtn} becomes
\begin{equation}
F^{m}_{i}=\frac{gf^{i}_{o}-f^{i}_{e}}{gf^{i}_{o}+f^{i}_{e}}.
\label{gainonly}
\end{equation}
This form assumes that there are no other effects, such as
changing sky transparency (which can become important for very large
or small values of $g$ when the flux in only {\it one} ray is
effectively being measured) or significant phase shift induced by the
half-wavelength retarder plate.  For $g \neq 1$ the observed intensity
changes at each retarder plate angle, depending on the flux and gain
for each ray. The ideal values of $f_{o}$ and $f_{e}$ (without
instrumental effects) are given as
\begin{equation}
f^{0}_{o}=\frac{1}{2}\overline{I}\left(1+q\right)=f^{2}_{e} \; ; \; f^{0}_{e}=\frac{1}{2}\overline{I}\left(1-q\right)=f^{2}_{o}
\label{fzero}
\end{equation}
\begin{equation}
f^{1}_{o}=\frac{1}{2}\overline{I}\left(1+u\right)=f^{3}_{e} \; ; \; f^{1}_{e}=\frac{1}{2}\overline{I}\left(1-u\right)=f^{3}_{o}
\label{fone}
\end{equation}
where the ideal intensity $\overline{I}=f_{o}+f_{e}$.  Using Eqns. \ref{gainonly}, \ref{fzero} and \ref{fone} in Eqns. \ref{eqeqtn} and \ref{eueqtn} gives
\begin{equation}
\epsilon_{Q}=
\frac{\left(g^{2}-1\right)\left(1-q^{2}\right)}{4g+
\left(g-1 \right)^{2}( 1 - q^{2})} \; ; \; \epsilon_{U}=
\frac{\left(g^{2}-1\right)\left(1-u^{2}\right)}{4g+
\left( g-1 \right)^{2}( 1 - u^{2})}.
\label{epsqu}
\end{equation}
These equations show that the instrumental signature corrections are dependent on the ratio of the gains of the two rays, the total degree of polarization (see Fig. \ref{fig:model:anal}) and the polarization angle.  In the ideal case, for $g=1$, Eqns. \ref{epsqu} give $\epsilon_{Q}=\epsilon_{U}=0$.  The sign of the corrections are dependent on the gain ratio, with
\begin{equation}
\epsilon_{Q}(g)-\epsilon_{U}(g)=-\left(\epsilon_{Q}(1/g)-\epsilon_{U}(1/g)\right).
\end{equation}
The magnitude of $\epsilon_{Q}-\epsilon_{U}$ increases for $g>1$, but reaches a stationary point at $g=3.85$.  For $g>3.85$ the magnitude of $\epsilon_{Q}-\epsilon_{U}$ decreases due to the predominance of the signal from one ray over the other.  This model, therefore, has limited application for realistic observing conditions, for variable observing conditions between observations at different retarder plate angles.  For $0.5<g<2$ and $p \lesssim 20\%$ the magnitude of this deviation, even when only one Stokes parameter is non-zero, is sufficiently small to be within
reasonable measurement uncertainties, while the assumption that
$\epsilon_{Q}-\epsilon_{U}=0$ can be used to identify if high polarizations are indeed present.
The dependence of $\epsilon_{Q}-\epsilon_{U}$ on the polarization angle is shown as Fig. \ref{fig:model:angle}.  The largest differences are measured for either pure $q$ or $u$ components, with mixtures of $q$ and $u$ leading to smaller differences.  The instrumental signature corrections are identical, regardless of total polarization, for $q=u$, $\theta=22$\fdg$5$.
The difference of the instrumental signature corrections is periodic over 90\degr.  For constant polarization and gain, the dependence on polarization angle is characterized by
\begin{equation}
\epsilon_{Q}-\epsilon_{U}=\epsilon_{Q}(0\degr,p,g)\cos (4\theta)
\end{equation}
where $\epsilon_{Q}(0\degr,p,g)$ is the $Q$ instrumental signature
correction for $\theta=0$\degr\, and $q=p$.

\subsection{Monte-Carlo Simulations} 
\label{section:models:monte}
\begin{figure}
\includegraphics[width=7cm, angle=270]{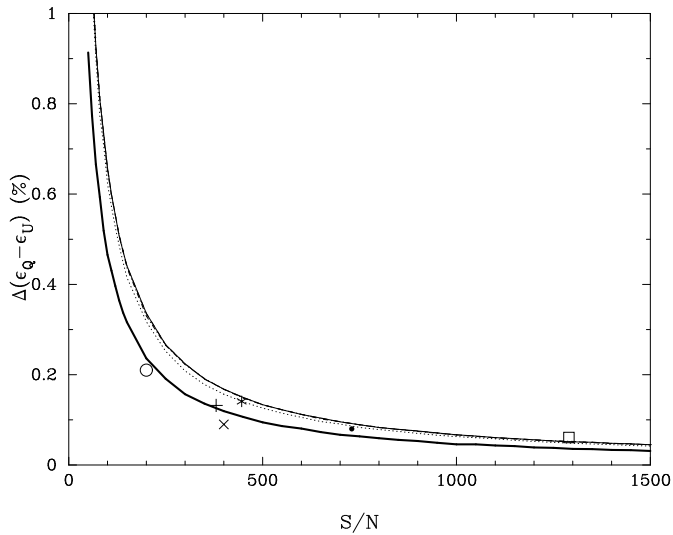}
\caption{The uncertainty (\%) on the difference between the
corrections $\Delta(\epsilon_{Q}-\epsilon_{U})$, of the $Q$ and $U$
Stokes parameters, as a function of the S/N.  Calculations were
conducted assuming the same parameters as FORS1 \citep{forsman}, with
$g = 1.1$, between the {\it o} and {\it e} rays.  All simulations were
conducted with $\theta=60\degr$.  Simulations were conducted with
p=1\% ({\it light solid}), p=5\% ({\it dashed}), p=10\% ({\it
dot-dashed}) and p=50\% ({\it dotted}).  The error on the measured $Q$
Stokes parameters is shown as the heavy black line.  Overlaid are
measured values of $\Delta(\epsilon_{Q}-\epsilon_{U})$ for six VLT
FORS1 observations of SN 2001ig at 03 Jan 2002 ($\Box$), HD 10038 at
07 Jul 2003 ($\bullet$), Vela 1 95 at 01 May 2005 ($+$), SN 2005hk at
23 Nov 2005 ($*$) and SN 2006X at 09 Feb 2006 ($\circ$) and 12 Feb
2006 ($\times$).}
\label{fig:model:varsn}
\end{figure}

\begin{figure}
\includegraphics[width=7cm, angle=270]{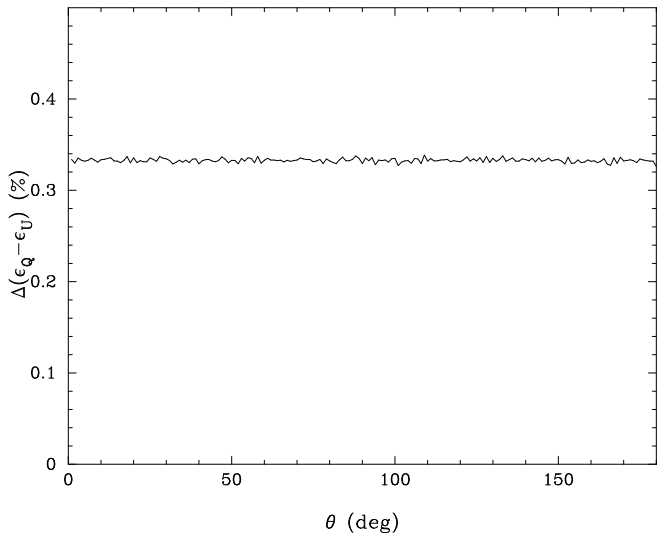}
\caption{The uncertainty (\%) of $\epsilon_{Q}-\epsilon_{U}$ as a function of the polarization
angle.  The calculation was made for $S/N=200$ and $p=1\%$.  The
absolute value of $\epsilon_{Q}-\epsilon_{U}$ is a factor of 30 lower
than the value of $\Delta(\epsilon_{Q}-\epsilon_{U})$.}
\label{fig:model:varchi}
\end{figure}

In order to test the importance of various realistic instrumental
effects, a Monte-Carlo model of a dual beam spectropolarimeter was
constructed in a similar style to \citet{2006PASP..118..146P}.  This
simulates the observation of the individual {\it o} and {\it e} rays
that would be observed for specific values of the total polarization
$p$ and the polarization angle $\theta$ for a given a value of S/N.
Different levels of gain and read-noise were applied to the $o$ and
$e$ rays, in order to simulate the contribution of these factors to
the corrections.  The polarization induced by optical components
preceding the analyzer, such as the telescope mirror, and polarization
due to a non-ideal Wollaston prism were not included.  For each
simulation, the value of the difference of the $Q$ and $U$ corrections
($\epsilon_{Q}-\epsilon_{U}$) was measured.  The standard deviation of
these values ($\Delta(\epsilon_{Q}-\epsilon_{U})$) about the mean
value of $\epsilon_{Q}-\epsilon_{U}$ was used as a measure of the
relative error.  These terms permit the problem to be studied in terms
of absolute deviations and the measurement error associated with the
corrections.\\ For simple signal dominated observations, approaching
ideal unvarying conditions, the behaviour of
$\epsilon_{Q}-\epsilon_{U}$ is consistent with the analytical form
presented in \S\ref{section:models:analytic}.\\ If there is a
significant gain difference between the $o$ and $e$ rays, such that
the ratio is $\gtrsim10$, the determined Stokes parameters are
dominated by signal measured in the ray with the larger gain.
Importantly, the normalized flux difference is no longer applicable,
and such observing conditions are unrealistic.  Similarly, significant
changes in the sky background ($\sim 100$) between observations at
each retarder plate angle can also induce deviations from
$\epsilon_{Q}=\epsilon_{U}$, but this is also primarily due to the
associated change in S/N between observations.\\ The value of
$\epsilon_{Q}-\epsilon_{U}$ was calculated for varying values of S/N.
Here the S/N is calculated at each retarder plate position.  In
general, it was found that $\epsilon_{Q}-\epsilon_{U}\approx 0$, with
$\Delta(\epsilon_{Q}-\epsilon_{U})$ dependent on the level of S/N in
the individual observations, as shown as Fig. \ref{fig:model:varsn}.
Since the determination of $\epsilon_{Q}-\epsilon_{U}$ depends on 8
individual measurements, the $o$ and $e$ rays at the four retarder
plate angles, the uncertainty $\Delta(\epsilon_{Q}-\epsilon_{U})$ is
$\sim\sqrt{2}$ larger than the uncertainty associated with the
individual Stokes parameter.  The quantity
$\Delta(\epsilon_{Q}-\epsilon_{U})$ is an estimate of the uncertainty
of a Stokes parameter calculated from incomplete observations.
Furthermore, the dependence of $\Delta(\epsilon_{Q}-\epsilon_{U})$ on
the polarization angle was also modelled, as shown in Fig. \ref{fig:model:varchi}.
As expected, the uncertainty on $\epsilon_{Q}-\epsilon_{U}$ is not directly
dependent on polarization angle, although $\epsilon_{Q}-\epsilon_{U}$
is itself offset.

\section{Observations using VLT FORS1}
\label{section:observations}
\begin{figure}
\includegraphics[width=9cm, angle=0]{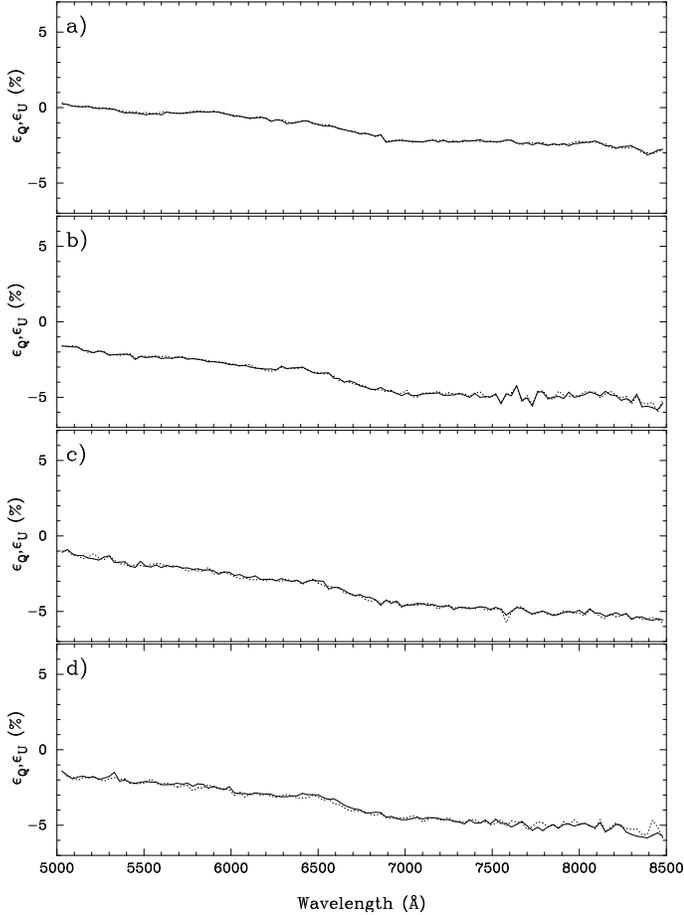}
\caption{$\epsilon_{Q}$ (solid line) and $\epsilon_{U}$ (dotted line),
as a function of wavelength, for four complete FORS1 observations of:
{\it a)} SN 2001ig at 03 Jan 2002 (S/N$\sim$1290), {\it b)}
unpolarized standard HD 10038 at 07 Jul 2003 (S/N$\sim730$), {\it c)}
polarized standard Vela 1 95 at 01 May 2005 (S/N$\sim$380), {\it d)} SN
2005hk at 23 Nov 2005 (S/N$\sim$446). S/N values are given at 6000\AA.
The data have be rebinned to 30\AA. }
\label{fig:obs:epanel}
\end{figure}

\begin{figure}
\includegraphics[width=6.5cm, angle=270]{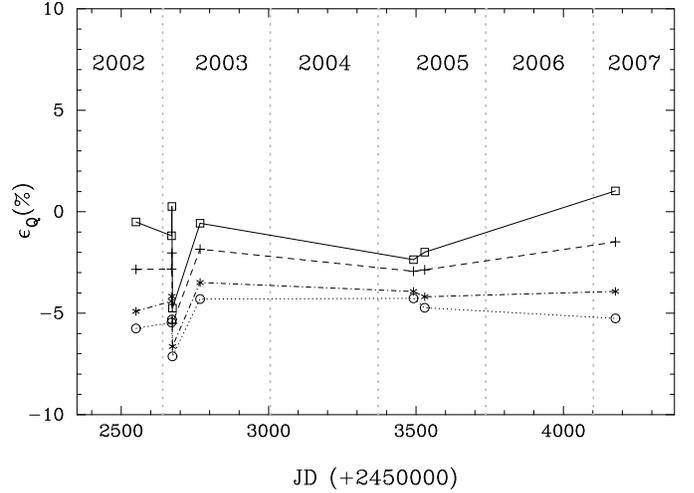}
\caption{The instrumental signature correction $\epsilon_{Q}$ for
unfiltered VLT FORS1 300V PMOS observations of Vela1 95 at 02 Oct
2002, 01 Feb 2003, 03 Feb 2003, 07 May 2003, 30 Apr 2005, 07 Jun 2005,
17 Mar 2007.  The data shown are for 1000\AA\ bins, centred on
5000\AA\ ($\Box$, solid line), 6000\AA\ ($+$, dashed line), 7000\AA\
($\ast$, dot-dashed line), 8000\AA\ ($\circ$, dotted line).}
\label{fig:obs:velaeps}
\end{figure}

\begin{figure}
\includegraphics[width=7cm, angle=270]{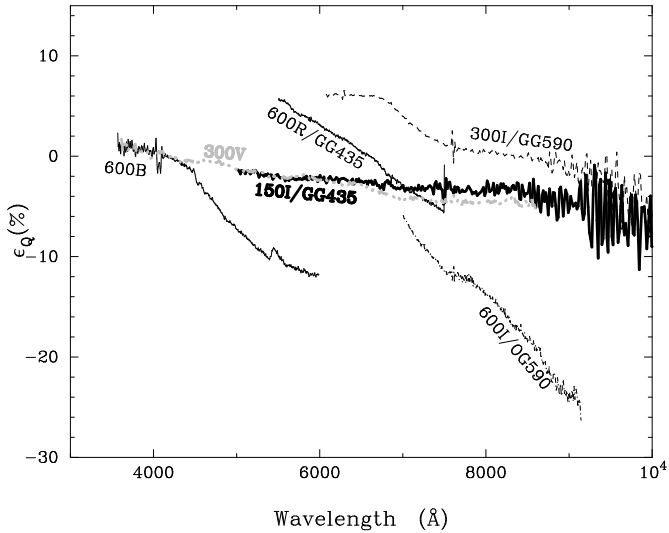}
\caption{The instrumental signature correction $\epsilon_{Q}$ for VLT FORS1 150I, 300I, 600B, 600R and 600I PMOS observations of polarized standard stars (listed in Table \ref{tab:obs:grism}).  The instrumental signature correction for the 300V grism is shown as the grey line.}
\label{fig:obs:grism}
\end{figure}

Spectropolarimetry observations of two SNe, one polarized standard
star and one unpolarized standard star, acquired using the ESO VLT
FORS1 instrument, were retrieved from the ESO
archive\footnote{http://archive.eso.org}, to determine if, under real
observing circumstances, $\epsilon_{Q}\approx\epsilon_{U}$.  These
observations were of SN 2001ig at 03 Jan 2002 \citep{my2001ig}, the
unpolarized standard HD 10038 at 07 Jul 2003 (previously unpublished),
the polarized standard Vela 1 95 at 01 May 2005 \citep{maund05bf}, and
SN 2005hk at 23 Nov 2005 \citep{my2005hk}\footnote{The reader is
directed to the referenced publications for the exact nature of the
polarization of each of these objects.}.  These observations were
conducted using the 300V grism, and were re-binned to 30\AA.  The
observations were complete, with data at N=4 retarder plate angles,
and were reduced in the standard manner as outlined by
\citet{maund05bf}.\\ Observations of SNe were specifically chosen as:
a) they represent the class of quick varying transients for which such
a technique might be useful; and b) despite having diverse
spectroscopic and polarimetric properties there has been a long term
VLT FORS1 program since 2000 to observe these events using similar
instrumental setups (namely the 300V grism).\\ As each of these
observations was complete, both $\epsilon_{Q}$ and $\epsilon_{U}$
could be determined; these are shown, as a function of wavelength, as
Fig. \ref{fig:obs:epanel}.  The corrections are observed to differ
slightly between epochs.  This implies that, just as new bias and flat
calibration frames are required at each epoch, the instrumental
signature correction, if it is to be applied to determine one Stokes
parameter, needs to be determined from data acquired at the same
epoch.  The corrections are wavelength dependent, but not observed to
be dependent on either the polarization or the spectroscopic
properties of these objects.  In instances of higher S/N (such as
Fig. \ref{fig:obs:epanel}a) the corrections are observed to be a
smooth function of wavelength, with more variability observed for
lower S/N observations.
Importantly, the largest deviations from $\epsilon_{Q}-\epsilon_{U}=0$
are observed for $\lambda > \mathrm{7000\AA}$, where there are a
number of telluric absorption bands and the response of the FORS1
instrument is falling.  For each of the four observations
$\Delta(\epsilon_{Q}-\epsilon_{U})$ was calculated over the wavelength
range 5500-6500\AA, to be representative of
$\Delta(\epsilon_{Q}-\epsilon_{U})$ at 6000\AA\ at which the S/N is
quoted.  For the given S/N at 6000\AA\ these values of
$\Delta(\epsilon_{Q}-\epsilon_{U})$ are plotted on
Fig. \ref{fig:model:varsn} in comparison with the Monte Carlo model.
The observed data agree well with the modelled dependence of
$\Delta(\epsilon_{Q}-\epsilon_{U})$ on the S/N and polarization of the
observations.\\ The observation of SN 2005hk (23 Nov 2005;
Fig. \ref{fig:obs:epanel}d) highlights the potential pitfalls of not
observing a complete dataset.  In the region 6500-7000\AA\ there are a
series of bad pixels which affected the $o$ ray of the observation
with the retarder plate at 45\degr, which could not be removed by bias
subtraction or correction with a normalized flatfield.  None of the
observations at other retarder plate angles, for either the $o$ and
$e$ ray, were affected.  In this case, therefore, the calculation of
$U$ by assuming $\epsilon_{U}=\epsilon_{Q}$ would lead to a serious
error in the Stokes parameter without a substantial decrease in the
S/N.  The quality of the reduction at any wavelength, for a given S/N,
can be directly measured by comparison of the $Q$ and $U$ instrumental
signature corrections, and the presence of such defects identified.
Importantly, the S/N is independent of the actual polarization, rather
it is dependent on the total flux in both the {\it o} and {\it e}
rays.  The reduction may, therefore, be questioned if the absolute
measured difference of $\epsilon_{Q}$ and $\epsilon_{U}$,
$|\epsilon_{Q}-\epsilon_{U}|_{m}$, exceeds some multiple of the
theoretically expected scatter, at a given S/N,
e.g. $|\epsilon_{Q}-\epsilon_{U}|_{m} >
3\Delta(\epsilon_{Q}-\epsilon_{U})|_{S/N}$.\\ In order to test the
temporal stability of the absolute instrumental signature corrections,
as a function of wavelength, a set of seven observations of the
polarized standard star Vela 1 95 were retrieved from the ESO archive.
These observations were acquired on 02 Oct 2002, 01 Feb 2003, 03 Feb
2003, 07 May 2003, 30 Apr 2005, 07 Jun 2005 and 17 Mar 2007, using an
{\it identical} instrumental setup: the 300V grism, no order
separation filter and $2048\times500$ windowing of the CCD detector.
There was not expected to be any significant effect due to detector
windowing, but for consistency between the compared observations the
windowing constraint was enforced.  In order to directly compare the
absolute values of the instrumental signature corrections at different
epochs, the data were rebinned to $1000\AA$\ centred on 5000,
6000, 7000 and 8000\AA.  The values for the instrumental signature
correction at these epochs are shown as Fig. \ref{fig:obs:velaeps}.
The absolute values of the instrumental signature corrections vary
between epochs.  Additionally the wavelength dependence of the
instrumental signature correction is also variable, with the gradient
of the correction across the wavelength range (corresponding to the
vertical distance between different wavelength points at the same
epoch, in Fig. \ref{fig:obs:velaeps}) changing.  This confirms that
instrumental signature corrections measured at different epochs cannot
be used to reduce incomplete data acquired at other epochs and, hence,
a minimum of N=3 observations are required.\\ The observed variability
in the instrumental corrections may arise from evolution of the
flatfield and of bias calibration frames, especially if the observed
object spectra are not at the exact same location on the detector for
each observation.  Wavelength-dependent slit losses due to
observations with the slits oriented with $PA=0\degr$, rather than the
parallactic angle, may contribute to the varying gradients of the
corrections.  Importantly, despite the variability, $\epsilon_{Q} =
\epsilon_{U}$ at each epoch.\\
The forms of the absolute values of $\epsilon_{Q}$ and $\epsilon_{U}$
are also dependent on the choice of grism.  For the VLT FORS1, linear
spectropolarimetry has been conducted with six grisms, for which
observations of representative standard stars were selected (as listed
in Table \ref{tab:obs:grism}).  The absolute forms of the corrections
for observations with all six grisms used for spectropolarimetry are
shown in Fig. \ref{fig:obs:grism}.  The 300V grim is used with
significantly higher frequency than the other grisms, such that the
stability of the instrumental signature corrections for the other
grisms cannot be tested with sufficient cadence or over a suitably
long time frame as was done for 300V above.  The principal
difference between the values of $\epsilon_{Q}$ and $\epsilon_{U}$
measured with different grisms is likely to arise from the response
function of the detector for given wavelength dispersions.  It is
important that the values of $\epsilon_{Q}$ and $\epsilon_{U}$ for any
particular grism cannot, in general, be extrapolated from those values
measured with other grisms (except, potentially, in the cases of the
300V and 150I grisms).

\begin{table}
\caption{\label{tab:obs:grism}Linear spectropolarimetry data, of standard polarized stars, for selected VLT FORS1 grisms.}
\begin{tabular}{llccc}
Grism& Object        & Date        & Program ID & Filters     \\
\hline
150I & Hiltner 652   & 19 Sep 1999 & 63.P-0002  & GG435\\
300V & Hiltner 652   & 05 Apr 2002 & 60.A-9203  & $\ldots$ \\
300I & BD 14 4922    & 23 Aug 2002 & 60.A-9203  & OG590     \\
600B & HD 126593     & 22 Aug 1999 & 63.P-0074  & $\ldots$  \\
600R & HD 126593     & 22 Aug 1999 & 63.P-0074  & GG435  \\
600I & HD 126593     & 22 Aug 1999 & 63.P-0074  & OG590   \\
\end{tabular}
\end{table}

\section{Discussion and Conclusions}
\label{section:discussion}
For spectropolarimetry where observations at N=3 retarder plate
positions are available, the assumption that
$\epsilon_{Q}=\epsilon_{U}$ can be employed to determine the
additional Stokes parameter.  This is a general property of dual
beam-spectropolarimeters, which have negligible intrinsic
instrumental polarization, and has been confirmed for VLT FORS1.  If the Wollaston prism is non-ideal, the
assumption of $\epsilon_{Q}=\epsilon_{U}$ is not true.  This technique is still
limited, as at least N=3 exposures in the standard observing sequence
are required, although at a moderate expense of precision.  Because
the corrections are dependent on calibrations applied to
the object data, the corrections calculated at different
epochs are not strictly identical; although, in some cases they have
been observed to be similar.  The assumption
$\epsilon_{Q}=\epsilon_{U}$ is not strictly true for observations of
objects with linear polarization $\gtrsim 20\%$.  The uncertainty on
$\epsilon_{Q}-\epsilon_{U}$ is dependent on the S/N of the
observations at each retarder plate angle and {\it not} the
polarization of the object being observed.  In some cases, and in the
particular cases of some SNe, this will permit reasonable
spectropolarimetry in instances when it was not possible to acquire
the complete set of observations.\\ In addition, the
$\epsilon_{Q}-\epsilon_{U}$ parameter can also be used to identify
defects and spurious polarization signatures in complete datasets.
The $\Delta(\epsilon_{Q}-\epsilon_{U})$ parameter can also be used to
determine the quality of the data reduction procedure as applied to
such datasets, as it is directly dependent on the S/N, and significant
deviations from $\epsilon_{Q}-\epsilon_{U}=0$ would indicate the presence of defects in the data.\\
The instrumental signature corrections are dependent on wavelength, and
are observed to be variable with time. Just as bias, flatfield and
wavelength calibration observations  are acquired for each
epoch, the instrumental signature corrections should be
determined at each epoch for each instrumental setup.  Fortunately,
this ``additional'' calibration information is derived from direct
observations of target objects, requiring no additional observation
time. The forms of $\epsilon_{Q}$ and $\epsilon_{U}$ are dependent on the choice of grism and the resulting dispersion across the detector.\\
Ultimately, the assumption of $\epsilon_{Q}-\epsilon_{U}=0$ is less
than ideal, leading to larger degrees of uncertainty in the extra
Stokes parameters by a factors of $\sim \sqrt{2}$ over parameters determined with N=4 observations.  The use of these corrections does permit, however, the utilisation of data which would, ordinarily, remain unused \citep{maund06aj}.

\begin{acknowledgements}
Based on observations made with ESO Telescopes at the Paranal
Observatories under programmes 60.A-9203(A), 63.P-0002, 63.P-0074,
68.D-0571(A), 76.D-0177(A) and
76.D-178(A).  The research of JRM is supported in part by NSF grant
AST-0406740 and NASA grant NNG04GL00G.  JRM thanks J. Craig Wheeler,
Ferdinando Patat and Dietrich Baade for their useful comments, and
also thanks Stefano Bagnulo, the referee, for his important
suggestions.
\end{acknowledgements}
\bibliographystyle{aa}

\end{document}